\documentclass[
aps,%
8pt,%
final,%
notitlepage,%
oneside,%
twocolumn,%
nobibnotes,%
nofootinbib,%
superscriptaddress,%
noshowpacs,%
centertags]%
{revtex4}

\usepackage[dvipsnames]{color}

\def\be{\begin{equation}}
\def\ee{\end{equation}}
\def\ba{\begin{eqnarray}}
\def\ea{\end{eqnarray}}

\def\12{{1\over 2}}

\def\ltsima{$\; \buildrel < \over \sim \;$}
\def\simlt{\lower.5ex\hbox{\ltsima}}
\def\gtsima{$\; \buildrel > \over \sim \;$}
\def\simgt{\lower.5ex\hbox{\gtsima}}

\begin{document}
 %\selectlanguage{english}
\title{The interaction of dark matter cusp\\ with the baryon component in disk galaxies}
\author{S.~A. Khoperskov}
\email{khoperskov@inasan.ru}
\affiliation{Institute of astronomy Russian academy of sciences, Moscow, 119017, Russia\\}
\author{B.~M. Shustov}
\affiliation{Institute of astronomy Russian academy of sciences, Moscow, 119017, Russia\\}
\author{A.~V. Khoperskov}
\affiliation{Volgograd state university, Volgograd, 119017, Russia\\}

\begin{abstract}
In this paper we examine the effect of the formation and evolution of the disk galaxy on the distribution of dark halo matter. We have made simulations of isolated dark matter (DM) halo and two component (DM + baryons). N-body technique was used for stellar and DM particles and TVD MUSCL scheme for gas-dynamic simulations. The simulations include the processes of star formation, stellar feedback, heating and cooling of the interstellar medium. The results of numerical experiments with high spatial resolution let us to conclude in two main findings. First, accounting of star formation and supernova feedback resolves the so-called problem of cusp in distribution of dark matter predicted by cosmological simulations. Second, the interaction of dark matter with dynamic substructures of stellar and gaseous galactic disk (e.g., spiral waves, bar) has an impact on the shape of the dark halo. In particular, the in-plane distribution of dark matter is more symmetric in runs, where the baryonic component was taken into account.
\end{abstract}

   \maketitle

\section{Introduction}

Recent progress in understanding of the formation and early
evolution of galaxies allows to construct a fairly clear picture of these processes
\cite{Steinmetz2003, Mo2010, ShustovKabanov2012}. According to current cosmological $\Lambda$CDM-models, gravitationally bound objects appeared from density fluctuations, which evolved hierarchically, i.e. in the <<bottom-up>> assembly of cosmic structures. Lowest mass objects were formed first, and the more massive structures were formed by mergings and accretion of these objects. These structures mostly constructed from the dark matter (DM), which average density was approximately $6-10$ times greater than the density of baryonic component. As a result, a large number of gravitationally bound objects were formed in a wide range of masses. Baryonic matter gradually settled in DM potential wells. First stars (population III) and protogalaxies were formed due accumulation of baryonic gas. First stellar population has changed the chemical composition of the primordial gas, enriching it in metals.

Compactness of the universe at early stages of evolution ($z = 10 - 30$) caused numerous mergers of DM-halos with each other, forming the large-scale structures. Approximately at $z = 1 - 10$ galaxies were formed. Merger of galaxies with similar sizes and masses (large merging) occurred mostly until  the epoch of $z \sim 1 - 2$. About same time galaxies acquired most of the angular momentum.
Further the galaxies, if they were not in a dense cluster, could have  only small merging, i.e. merging with satellites of masses less than 10\% of the host galaxy. However, despite the apparent success of the construction of adequate models of the formation and evolution of galaxies, many issues remain unresolved. The essence of these questions is the discrepancy between the theoretical conclusions of models and observable characteristics of galaxies.
Here are some of the problems typical for cosmological models of formation and evolution of disk galaxies.

\begin{itemize}
  \item The density distribution of dark matter in the central regions of galaxies according to the calculations is much more than ocute the observable one (the problem of the central cusp)~\cite{Doroshkevich2012}.
  \item The correlation  between mass and size in models of galaxies is less pronounced than observations show~\cite{Dutton2007}.
  \item In the models,  galaxies are generally not formed without a central bulge~\cite{ZasovSilchenkoUFN}.
  \item Galaxies in the models only partially are balanced by the rotation~\cite{Roskar2010}.
  \item <<Missing galaxy problem>>: DM simulations predict more subhaloes (low-mass galaxies in groups) than number of dwarf galaxies observed~\cite{Governato2007}.
\end{itemize}
These problems are particulary linked to insufficient numerical resolution: typical spatial scale of the baryonic matter evolution is of many orders of magnitude smaller than the spatial scale of dark halo evolution ($\sim 10^2 - 10^4$ kpc). Even for the galactic disk ($\sim 10$ kpc) there is no comprehensive numerical methods, adequate including by star formation processes, chemical enrichment of the interstellar medium, supernova explosions, heating and cooling of the gas, e.t.c, because the spatial scales of these processes usually do not exceed few parsecs.
Therefore all models have to work in the approximation of <<subgrid physics>>. In any case, increasing of the spatial resolution and accurate treatment of stellar-gaseous component in the models is an important direction to improve the theoretical approach.

In this paper we considered the influence of the formation and evolution of the disk galaxies on the dark matter halo distribution. A particular emphasis was made on the central cusp problem and on the baryonic matter influence on the shape of a dark halo, during the disk galaxy formation and evolution.

The problem of the central cusp is well known from cosmological simulations of the evolution of dark matter and the formation of baryonic structures. The problem is that in numerical models the volume density $\rho$ in the center of the DM halo tends to singularity, forming the so-called central cusp. But in the observations such cusps of the density distribution are not detected.
The density profile in the DM-halo is described by the approximation $\rho \propto r^\alpha$ where $-1.5 < \alpha < -1$. For example, for the most commonly used profile NFW \cite{Navarro1996}:

\begin{equation}\label{eq::NFW}
\displaystyle \rho \propto \frac{\rho_0}{ (r/r_s)(1 + (r/r_s)^2)}\,,
\end{equation}
 where $r$ is the distance from the halo center, $\rho_0$ and $r_s$ are the parameters which describe a particular halo, $\alpha = -1$.

Observed density distributions of stars correspond to the distributions of density in the DM-halo with $\alpha > -1$ («core» type distribution). For comparison with observations they commonly use data on the rotation curves (or velocity dispersion) in the central regions of dwarf galaxies  \cite{deBlok2008}. Dwarf galaxies are more appropriate objects to determine the  distribution of the density of dark halo, which can dominate even in the central regions of these galaxies \cite{Gentile2005}. For more massive galaxies, observations also confirm the «core» type density distribution of dark halo. For example, in \cite{Spano2008}  detailed rotation curves were recovered, based on $Í\alpha$ kinematic survey of 36 close spiral galaxies. The results of decomposition of rotation curves of these galaxies  argue in favor of quasi-isothermal density distribution of the dark halo (i.e. $\alpha$ is closer to $0$ than to $-1$ as in the NFW distribution).

Definitely many researchers tried to find the solution of the cusp problem. Two approaches can be considered. According to first approach the cusp problem is treated  taking into account the properties of dark matter. For example, in \cite{Doroshkevich2012,Mikheeva2007} it was proved that the cosmological random motions of «heated» DM particles in collapsing proto-halos. Lead to the suppression of the cusp-like density profiles inside the forming halo, promote the formation of DM-nucleus in galaxies and allow to explain the difference between observed and obtained in the numerical experiment rotation curves of galaxies. Analytical findings made in this approach, should be confirmed by the numerical N-body models, this is possible with the improving of the spatial resolution of the central regions of the halo. In \cite{Medvedev2010} CDM model is considered as an ensemble of elementary particles with mixture of flavours~(such as neutralinos; particles that may gradually «evaporate» from the potential well).

According to the second, more spread approach the influence of baryons, that have been accumulated in the dark matter potential well on the distribution and kinematics of dark matter is considered. Following $\Lambda$CDM paradigm, various authors \cite{Abadi2010,Khoperskov-etal-2010!z-str} have examined the «post»-evolution of dark matter under the influence of baryons. Baryonic matter impacts on the distribution of DM caused via supernovae and/or dynamic frictions and tidal effects. Of course, manifestation of these mechanisms become substantial $1 - 2$ billion years after the galaxy formation. In paper \cite{Weinberg2002} which the linear perturbation theory and N-body simulations were used in, it was shown that the bar in the galactic disk may transform the dark matter central cusp (which had formed earlier) into the «core» type distribution.

There are numerous models, indicating the absence of central symmetry in the dark matter distribution within the galactic disk \cite{Cooper-etal-2009!Galactic-stellar-haloes-CDM-model, Kuhlen-etal-2007!shapes-orientation-subhalo}.
Particularly in work \cite{Allgood-etal-2006!shape-halo-N-body} on the basis of cosmological
N-body models of galaxy formation, it was shown that the dark halos have a triaxial shape.
The typical ratio for the semi-axes according to the various data: in the plane of the disk $q=b/a\sim0.8-0.95$, in the perpendicular direction $s=c/a\sim0.6-0.7$, where $a$ is the semi-major axis. The parameters of the non-axisymmetry depend on redshift, relative mass of halo and other parameters of the model.
Observational manifestations of a triaxial dark matter distribution are given in several papers: distortion of the gas layer beyond the optical radius of the galaxy (warping) \cite{Dubinski-Chakrabarty-2009!Warps-Triaxial-Halo, Roskar2010}, non-spherical shape of the isophotes of X-ray emission of hot gas in the halo \cite{Das-etal-2010!dark-matter-X-ray}, dynamics of the outer galactic polar rings \cite{Whitmore-etal-1987!halo-polar-ring}, observing the kinematics of hyper velocity stars (HVS) \cite{Brown-Geller-Kenyon-2008!MMT-Hypervelocity-Star}, dynamics of tidal streams \cite{Belokurov-etal-2007!Hercules-Aquila-Cloud, Helmi-2004!Is-dark-halo-our-Galaxy-spherical}.

According to the models, the dark halo shape is more symmetrical in the center than at
the far periphery. Apparently, on the scale lengths of the galactic disk, dark and
baryonic matter masses are comparable and the influence of the baryonic component on the characteristics  of dark matter is the most significant. However, large-scale
$\Lambda$CDM models do not provide an opportunity to follow in detail the structure
 and transformation of the dark halo on the small scale lengths (within the optical
 radius of a galaxy) under the influence of the baryonic component.

%%%%%%%%%%%%%%%%%%%%%%%%%%%%%%%%%%%%%%%%%%%%%%%%%%%%%%%%%%%%%%%%%%%%%%
\begin{figure*}
%\setcaptionmargin{5mm}
%\renewcommand{\baselinestretch}{1}
%\onelinecaptionsfalse
\includegraphics[width=1\hsize]{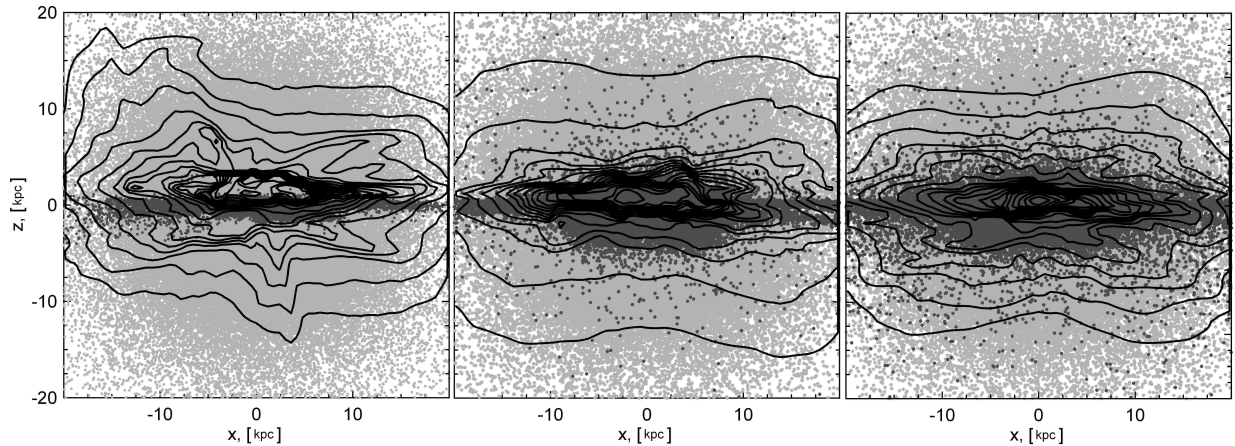}
\caption{The evolution of the baryons (dark gray dots - stellar particles, black outlines --- gas) in the gravitational potential of dark matter (light gray dots) at the times $t=10^9$, $t=6\cdot10^9$, $t=10\cdot10^9$ yr}
\end{figure*}\label{fig1evol}
%%%%%%%%%%%%%%%%%%%%%%%%%%%%%%%%%%%%%%%%%%%%%%%%%%%%%%%%%%%%%%%%%%%%

In this paper we have constructed a simple model of the formation and evolution of isolated galaxy in the gravitational field of dark halo taking into account the star formation, feedback (supernova explosions and the return of gas by normal stars) and the radiative cooling and heating of interstellar gas. We applied this model to study the possible role of baryons in the blurring of the central cusp of the dark halo density and transformation of the dark halo shape. In Section 2 the formulation of problem is given, an algorithm, software and input physics are described. In section 3 main results are presented. Section 4, contains a critical discussion of the results. The main findings are given in the conclusion (Section 5).

\section{Initial conditions and numerical methods}
We consider evolution of dark matter halo with and without baryonic component. The evolution of the galaxy, consisting of a dark halo and the baryonic component was considered at $10\cdot 10^{10}$~yr time scale.
At the initial moment, the matter contained within the computational domain is made up of dark halo ($10^{10}M_\odot$) and gas ($5\cdot 10^9 M_\odot$).
In the process of evolution, the gas can accrete on the galaxy from the outside and also leave the galaxy as a result of the evolutionary dynamic processes. At the initial moment gas with a temperature $T = 10^3$ K was evenly distributed over the halo, rotating at a rate of $0.5 \cdot V_c(r)$ ($V_c(r)$ is the circular velocity).
Typically, in theoretical models of formation and early evolution of galaxies, the potential of the dark halo is defined as an external factor with the given density profile of dark matter. In our work we started the calculation from the quasi-isothermal distribution of DM in order to investigate of the formation of cusp-distribution of dark matter density in the process of dynamic evolution of the DM-halo.

Numerical treatment of gas dynamics was made with TVD MUSCL third-order approximation numerical scheme in cartesian coordinate system \cite{vLeer1979}. To ensure conservatism of the sheme the finite-volume approximation of the variables was used. The minmod limiter was used \cite{Wada2001} in the calculations. The algorithm and software package based on it, were described in \cite{buma_khsa_Eremin-2010!VolGTU,Khoperskov-etal-2010!z-str} these were applied for solving problems of the galactic disk dynamics~\cite{Khoperskov-2012!Halo-gas-spiral, Khoperskov-Zasov-Tiurina-2003!GravitInstab}. Gas-dynamic runs were carried on a grid with spatial resolution of $70$ pc. The dynamic of dark matter particles ($N = 10^6$) and stars ($N = (0 — 0.25) \cdot 10^6$) was calculated using N-body technique. Total gravitational potential of the test star- and DM-particles and gas was calculated by TREEcode method \cite{Barnes1986}.

%%%%%%%%%%%%%%%%%%%%%%%%%%%%%%%%%%%%%%%%%%%%%%%%%%%%%%%%%%%%%%%%%%%%%
\begin{figure}
\includegraphics[width=1\hsize]{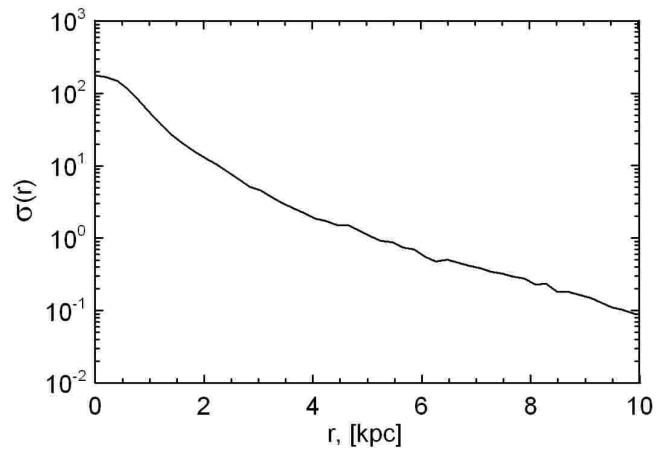}
\caption{The radial distribution of surface density of the stellar disk ($t=10\cdot10^9$ yr).}
\end{figure}\label{fig2dr}
%%%%%%%%%%%%%%%%%%%%%%%%%%%%%%%%%%%%%%%%%%%%%%%%%%%%%%%%%%%%%%%%%%%%

The boundary of computational domain are open. This enable free exchange (accretion and/or outflow) of baryons, which plays an important role in the process of galaxy evolution. The radiative cooling processes were taken into account, their rate was determined according to \cite{Katz1992}.

The process of star formation was simulated as follows.
At each step of integration cells were selected, criteria for selection were: the surface density of gas $\sigma > 10^4$~Ì$\odot$/pc$^2$ and the gas temperature $T < 50$ K. If the dynamical time in the cell $\triangle t_{dyn} = (3\pi/32G\rho)^{1/2}$ is lower than the integration step $\triangle t_{int}$, then the test star particle was placed into the cell. This particle can be represented as a star cluster~($10^3-10^4$ massive stars). We calculated the mass of such star particle:
\begin{equation}\label{eq::m_st}
    m^* = \xi_{SF} \cdot \rho_{gas} V_{cell}\,,
\end{equation}
where $\xi_{SF}$ is the star formation efficiency, $\rho_{gas}$ is the gas density, $V_{cell}$ is the volume of one cell. However, we did not consider the initial mass function of stars in such star particle (star cluster). The values of the efficiency of star formation  $\xi_{SF}$ were 1\% and 5\%. The initial velocity of the test particle corresponded to the gas velocity, from which it was formed. Kinematics of such particles was calculated taking into account the gravitational potential of the gas, dark halo and other stellar particles.

Star clusters (test particles) during the calculation injected mass and energy into the galaxy due to two mechanisms: the explosions of supernovae and stellar winds. Different approaches to implementation of a supernova explosion in the interstellar medium are used by various authors: the injection of thermal energy \cite{Katz1992, Mori1997, Rosen1995} or injection of the kinetic  energy of the supernova remnant~\cite{Navarro1993, Wiebe1998}. In our algorithm, we applied  the first approach. At each step of the integration ($\triangle t_{int} \sim 10^3 - 10^4$~yr) we tested the possibility of a supernova explosion in the cluster. The probability of such event is $R \triangle t_{int} m^*$, where  $R$ is the number of supernova explosions per $1 M_\odot$ per year. For Milky Way typical value is $R=10^{-13}$~$Ì_\odot^{-1}$~yr$^{-1}$, for starburst galaxies, $R$ reaches $10^{-11}$~$M_\odot^{-1}$~yr$^{-1}$. We have used a control parameter $E_{SN}$ in order to vary the contribution of supernovae in these range of values of $R$. For $E_{SN}$ were used values  $0.1$ and $0.2$. When a supernova with the energy $10^{51}$~erg explodes in the star cluster, this energy is partially transferred to the interstellar medium. Part of this energy is converted into the kinetic energy. A second mechanism for the transfer of matter from test particles to gas is the mass loss of stars. Return of gas into the ISM depends on stellar mass and could be significant (for example see \cite{Igumenshchev1990}). We assumed that at each step of integration of the system of equations describing the dynamics of star clusters, they lose some mass, so that each particle in $10 ^{10}$~ years returns to the gas half of its original mass.

\section{Formation and evolution of isolated galaxy in the field of dark halo}

The evolution of baryons in the gravitational potential of dark matter shown in fig.~ 1.
Initially gas concentrates to the midplane, where it cools down and conditions for star formation are. Herewith a thin galactic disk forms rapidly
($\sim 10^7-10^8$ yr).
Due the gravitational instability and the dynamic interaction with the DM- particles
galactic disc heats up and becomes much thicker.
In the later stages of evolution the central stellar bulge is formed, this indicate that the formation of the disc is going on <<outside-in>>.
This can caused by due to the gaseous streams from the external region,
stimulating star formation in the outer parts of the disk (see below).

Stellar disk, has a typical exponental profile of the surface density $\sigma(r) \sim \exp(—r/L)$ (see fig.~2). Furthermore, the stellar component density distribution demonstrate the presence of structure. In fig.~3 we show the relative perturbation of the surface density $\Sigma(r,\varphi)$:

%%%%%%%%%%%%%%%%%%%%%%%%%%%%%%%%%%%%%%%%%%%%%%%%%%%%%%%%%%%%%%%%%%%%%
\begin{figure}
\includegraphics[width=0.8\hsize]{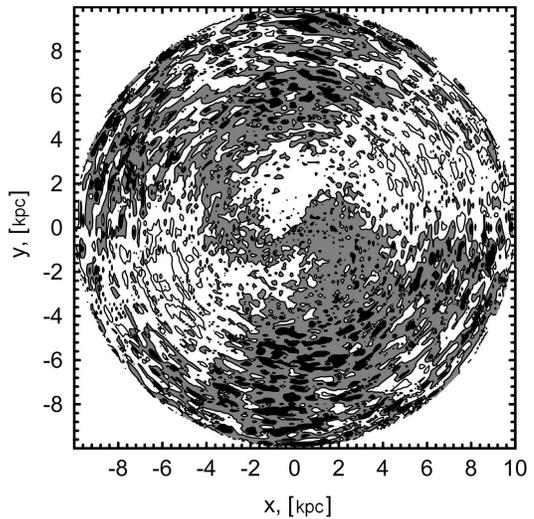}
\caption{Relative perturbation of the surface density $\Sigma(r,\varphi)$ of the stellar disk  ($t=10\cdot10^9$ yr).}
\end{figure}\label{fig3value}
%%%%%%%%%%%%%%%%%%%%%%%%%%%%%%%%%%%%%%%%%%%%%%%%%%%%%%%%%%%%%%%%%%%%

\begin{equation}\label{eq::DisturbDens}
\Sigma(r,\varphi) = \frac{(\sigma(r,\varphi) - \langle
\sigma(r,\varphi) \rangle_{\varphi}} {\langle \sigma(r,\varphi)
\rangle_{\varphi}} \,,
\end{equation}
herein $\langle ...\rangle_\varphi$ is the averaging over the azimuthal angle. The global two-arm spiral structure with the amplitude about $1 - 3\%$ is notable. It could be formed under the action of gravitational instability (for example see \cite{Sellwood2011, Griv-2004!Jeans-unstable}) or an inhomogeneous distribution of dark matter \cite{Khoperskov-2012!Halo-gas-spiral, Khoperskov2010arxiv}.

The dynamics of the gas in the disk plane is shown in fig.~ 4. In addition to the regular picture of rotation, gas streams from external areas and the outflows are visible. The velocities of streams is comparable with the rotational speed, but the gas densities in these areas are smaller.

%%%%%%%%%%%%%%%%%%%%%%%%%%%%%%%%%%%%%%%%%%%%%%%%%%%%%%%%%%%%%%%%%%%%%
\begin{figure}
\includegraphics[width=0.8\hsize]{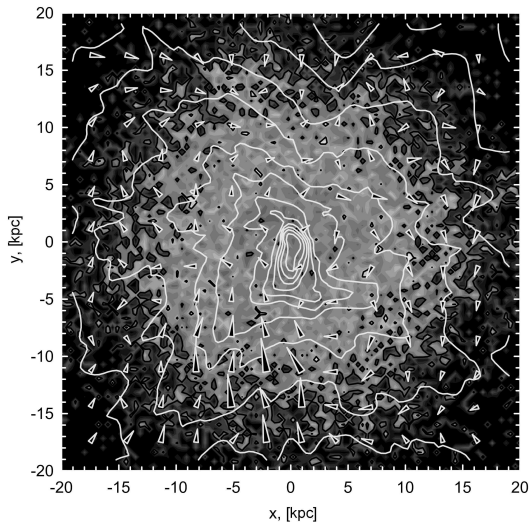}
\caption{The velocity field of the gas component in the plane of the galactic disk. White isolines show the surface density of the gas, gray --- the surface density of the stellar disk.}
\end{figure}\label{fig4filo}
%%%%%%%%%%%%%%%%%%%%%%%%%%%%%%%%%%%%%%%%%%%%%%%%%%%%%%%%%%%%%%%%%%%%

In fig.~ 5 (right panel) we show the profiles of volume DM density along the axis of rotation of the galaxy) for two models of the galaxy (with baryons and without baryons). In a model, which includes DM only,  the formation of a central density pike is observed.
There is a  difference in the radial dependence of the density profile of dark matter: in the calculations, taking into account the baryons, density behaves more smoothly in the central region (fig.~5 left panel).
We do not illustrate the course of evolution in detail, let's give only a brief description of the system evolution.
The baryonic gas coming into the galaxy is accumulated in the center.
If there were no star formation, the concentration of baryons in the center will increase the compression of dark matter.
However, as seen in fig.~ 5, the density distribution of DM in the presence of the baryonic component becomes «wider».
The reason for this is the star formation that begins in the dense central gas and leads to the appearance of supernovae, which are the powerful sources of energy and gas from star-forming region.
As is evident from fig.~ 5  the increase in value
$E_{SN}$ leads to more intense ejection of baryons from the central
region and accordingly to a flatter shape of DM distribution.
Ejection of gas leads to a redistribution of baryons in the center, which is similar to the «core»-type profile.
The gravitational contribution of the baryons in the central
region is comparable to the DM. As it'seen in fig.~5 the density of baryonic matter
in the center of the galaxy is $5-7$~times higher than density of the DM. So the distribution of DM is
also getting similar to the «core»-type distribution.
Therefore, our results suggest that the «normal»
evolution of the baryonic component of galaxies leads to the evolutionary
smoothing of the cusp, even if it was formed earlier, and prevents its
 occurrence in the later stages of evolution.
%%%%%%%%%%%%%%%%%%%%%%%%%%%%%%%%%%%%%%%%%%%%%%%%%%%%%%%%%%%%%%%%%%%%%
\begin{figure*}
\includegraphics[width=0.7\hsize]{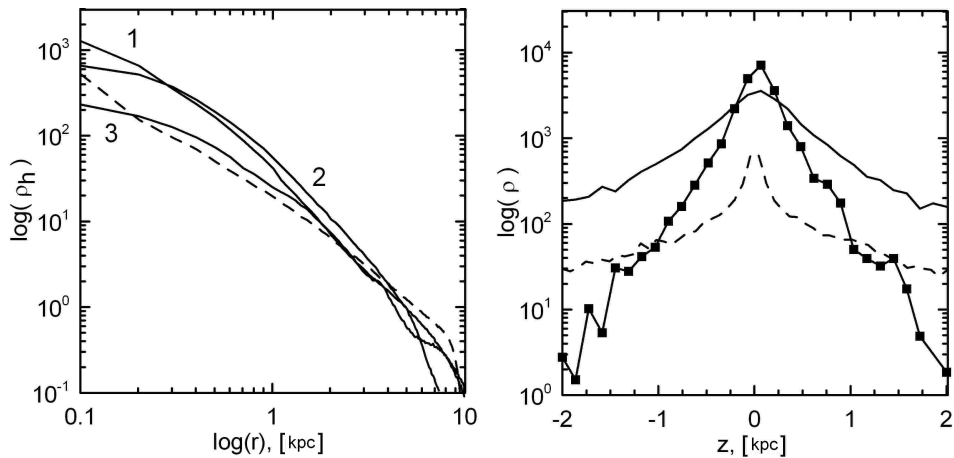}
\caption{The radial distribution of the average volume density of dark matter is shown on left panel: the dashed line --- in the model without baryons, the solid lines correspond to different models with baryonic component (1 --- $E_{SN}=0.1$, $E_{SF}=0.05$; 2 --- $E_{SN}=0.1$, $E_{SF}=0.01$; 3 --- $E_{SN}=0.2$, $E_{SF}=0.05$).
The distribution of volume density perpendicular the plane of the galaxy  is shown on left panel: the dashed line corresponds to the density of DM in the model without baryons, the solid line with squares --- the density distribution of baryonic matter,  the solid line ---  the density distribution of DM in the model, with the baryon component ($E_{SN}=0.1$, $E_{SF}=0.01$).}
\end{figure*}\label{fig5cusp}
%%%%%%%%%%%%%%%%%%%%%%%%%%%%%%%%%%%%%%%%%%%%%%%%%%%%%%%%%%%%%%%%%%%%%

Now consider the problem of «large-scale» influence of the baryonic component on the dark halo shape. The distribution of dark matter within the galactic disk is of significant interest, in terms of possible observational manifestations. Using our model, we examined the evolutionary changes of the shape of dark halo in the neighborhood of the galactic disk under the influence of baryons.

%%%%%%%%%%%%%%%%%%%%%%%%%%%%%%%%%%%%%%%%%%%%%%%%%%%%%%%%%%%%%%%%%%%%%%
\begin{figure}
\includegraphics[width=1\hsize]{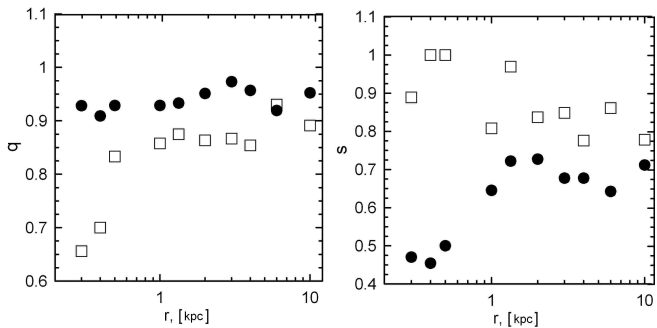}
\caption{Distribution of the parameters of non-axisymmetrical density of dark matter: open squares correspond to the model without baryons, dark circles --- the model with baryons. On the left --- the value of $q$ in the plane of the disk ,on the right $s$ --- perpendicular to the disk plane.}
\end{figure}\label{fig6axis}
%%%%%%%%%%%%%%%%%%%%%%%%%%%%%%%%%%%%%%%%%%%%%%%%%%%%%%%%%%%%%%%%%%%%

In fig.~6 the spatial distributions of the parameters $q$ and $s$ (the ratios of small semi-axis of the ellipsoid to the semi-major axis: in the galactic plane in the perpendicular direction, respectively) are shown.
In the absence of baryons, the deviation of the distribution of density of the DM from the circular one is insignificant on the periphery and quite noticable in the central regions. As it is evident from the distribution of $q$ in the model, which includes DM and baryons, the distribution of DM due to interaction with baryonic matter of the galaxy in the plane of the disk becomes more symmetric. It seems that the spiral structure and/or the bar arising in galactic disk, interacting with a dark halo leads to the decrease of non-axisymmetry~\cite{Abadi2010}. In the direction perpendicular to the disk plane, DM halo becomes more oblate, this is a consequence of the deeper baryonic
gravitational potential well, in which the dark matter accumulates.

\section{Discussion}
The initial mass of baryons in the model is $5\cdot10^9M_\odot$ at the $t = 10^{10}$ years it increased up to $10^{10} M_\odot$ due to the openness of the system changed. Despite the evolutionary increase of the mass, the galaxy formed i our model can  be classified as a dwarf. As already has been noted, the dwarf galaxies are the most suitable objects to determine the distribution of the dark matter density, which these systems is usually dominated. Anyway DM determines a significant share of the potential in the central regions of galaxies. Three reasons for this are considered.

First, from the shallow gravitational potential well  of dwarf galaxy, the gas leaves the galaxy more easily than from the nuclei of massive galaxies. In \cite{Shustov2000} it was shown that there is a critical mass of gas that creates a potential well, which for single supernova explosion is deep enough to prevent the ejection of a substantial amount of gas in the intergalactic medium (in the original paper proto-globular clusters studied, but those are similar the protodwarfs galaxy). The critical mass was $10^7 - 10^8 M_\odot$ it depends on the parameters (the initial density distribution, the details of the cooling law, etc). In our runs supernova explosions rate does not depend on time ($R=const$) that is not suitable for realistic galaxy evolution. In the more massive galaxies where $R$ can be the two orders higher than the current rate, a significant ejection of gas may also occur~(the mode of star formation burst).
The second reason is that the low-mass DM halo do not have time to accumulate a substantial amount of gas from their surroundings. If the hierarchical merging of subhalos occur, relatively low value of the densities ratio $\rho_b/ \rho_{DM}$ remain.
Only after the reaching of a certain size (mass) of galaxy, accretion of gas from the intergalactic medium becomes comparable to the efficiency accumulation of mass in the process of minihalos merging.
Third, the frequent collisions of many dwarf galaxies lead to the loss of gas from them, and this decreases the ratio $\rho_b/\rho_{DM}$.

The fact that  the effect of smoothing the cusp of the DM density expressed for our model galaxy, allows to assume that the effect of smoothing of the cusp under the influence of the evolution of the baryonic component in the central regions of such galaxies does occur. For more massive galaxies, the evolution in the central regions is mainly determined by baryons.

Quite recently papers appeared, confirming our conclusion on the causes of
 smoothing of the cusp. In works \cite{Governato2012,Pontzen2012} $\Lambda$CDM simulations of the early evolution of galaxies were carried out.
 Authors conclude that the activity of supernova does flatten the initially cusp-density
  distribution of DM.
%----------------------- Section 6 -------------------------------

\section{Conclusion}
 Evolutionary model of the formation and evolution of isolated disk galaxy was constructed on the time interval of $10$~billion years. The high spatial resolution in three-dimensional model has allowed to identify the role of small-scale processes (star formation and feedback) in the transformation of the dark halo shape under the influence of baryonic matter.
The mechanism of the smoothing of the density cusp in the distribution of dark matter in the central region of disk galaxies is proposed. The process of star formation, is accompanied by an outflow of matter from the central regions of galaxy. The cusp of DM density distribution is smoothed due to the gravitational influence of baryons, since the density of baryons in the galaxy's center is higher than the density of the DM.
The interaction of dark matter with the small-scale dynamical structures of the stellar disk (spiral density waves, star bar) leads to a symmetrization of dark matter halo in the galactic plane ($b/à \sim 0.9-0.95$).

%----------------------- Section 12 ------------------------------
\section{Acknowlegments}

\noindent

This work was supported by the grant of RFBR (11-02-12247-ofi-m-2011, 10-02-00231, 12-02-00685-a), the grant of President of Russian Federation (SS-3602.2012.2) and by the non-profit foundation <<Dynasty>>. Numerical calculations were conducted on the supercomputers <<Lomonosov>> and <<Chebyshev>> (SRCC MSU) with the support of A.V.~Zasov and N.V.~Turina.

%\end{multicols}

\begin{thebibliography}{99}
\bibitem{Steinmetz2003} M. Steinmetz, Astrophys. and Space Science, \textbf{284}, 325 (2003)
\bibitem{Mo2010} H. Mo, F. van der Bosch, S. White, Galaxy formation and evolution, Cambridge unversity press (2010)
\bibitem{ShustovKabanov2012} B.M.~Shustov, A.A.~Kabanov, Physics of Space: the 41st Annual Student Scientific Conference, Edited by P.E. Zakharova et al. Publ. The Ural Federal Univ. 200 (2012)
\bibitem{Doroshkevich2012} A. Doroshkevich, V. Lukash, E.A. Mikheeva, UFN, \textbf{182}, 3 (2012)
\bibitem{Dutton2007} A.A. Dutton, F.Ñ van den Bosch, A. Dekel and et al., Astrophys. J., \textbf{654}, 27 (2007)
\bibitem{ZasovSilchenkoUFN} A.V.~Zasov, O.K.~Sil'chenko, UFN, \textbf{180}, 434 (2010)
\bibitem{Roskar2010} R. Roskar, V.P. Debattista, A.M.Brooks et al., Mon. Not. R. Astron. Soc., \textbf{408}, 783 (2010)	
\bibitem{Governato2007} F. Governato, B. Willman, L. Mayer et al,  Mon. Not. Roy. Astron. Soc.,  \textbf{374}, 1479 (2007)
\bibitem{Navarro1996}  J.F. Navarro, C.S. Frenk, S.D.M. White, Astrophys. J., \textbf{463}, 563 (1996)
\bibitem{deBlok2008} W.J.G. de Blok, F. Walter, E. Brinks et al., Astron. J., \textbf{136}, 2648 (2008)
\bibitem{Gentile2005} G.~Gentile, A.~Burkert, P.~Salucci, and et. al., Astrophys. J. \textbf{634}, 145 (2005)
\bibitem{Spano2008} M. Spano, M. Marcelin, P. Amram et al., Monthly Not. Roy. Astron. Soc., \textbf{383}, 297 (2008)
\bibitem{Mikheeva2007} E. Mikheeva, A. Doroshkevich, V. Lukash., Nuovo Cimento  Serie, \textbf{122}, 1393 (2007)
\bibitem{Medvedev2010} M.V. Medvedev, J. Phys. A: Math. Theor., \textbf{43}, 372 (2010)
\bibitem{Abadi2010} M.G. Abadi, J.F. Navarro,  M. Fardal et al., Mon. Not. R. Astron. Soc., \textbf{407}, 435 (2010)
\bibitem{Weinberg2002} M.D. Weinberg, N. Katz, Astrophys. J., \textbf{580}, 627 (2002)
\bibitem{Cooper-etal-2009!Galactic-stellar-haloes-CDM-model} A.P. Cooper, S. Cole, C.S. Frenk et al., Mon. Not. Roy. Astron. Soc., \textbf{406}, 744 (2010)
\bibitem{Kuhlen-etal-2007!shapes-orientation-subhalo} M. Kuhlen, J. Diemand,  P. Madau, Astrophys. J., \textbf{671}, 1135 (2007)
\bibitem{Allgood-etal-2006!shape-halo-N-body} B. Allgood, R.A. Flores, J.R. Primack et al, Mon. Not. Roy. Astron. Soc., \textbf{367}, 1781 (2006)
\bibitem{Dubinski-Chakrabarty-2009!Warps-Triaxial-Halo}
 J. Dubinski, D. Chakrabarty , Astrophys. J., \textbf{703}, 2068 (2009)
\bibitem{Das-etal-2010!dark-matter-X-ray}
 P. Das,  O. Gerhard, E. Churazov et al., Mon. Not. R. Astron. Soc.,\textbf{409}, 1362 (2010)
\bibitem{Whitmore-etal-1987!halo-polar-ring}
 B.C. Whitmore, D.B. McElroy, F. Schweizer, Astrophys J., \textbf{314}, 439 (1987)
\bibitem{Brown-Geller-Kenyon-2008!MMT-Hypervelocity-Star}
 W.R. Brown,  M.J. Geller,  S.J. Kenyon, preprint(arXiv:0808.2469) (2008)   \bibitem{Belokurov-etal-2007!Hercules-Aquila-Cloud} V. Belokurov et al., Astrophys. J. Letters, \textbf{657}, 89 (2007)
\bibitem{Helmi-2004!Is-dark-halo-our-Galaxy-spherical}
 A. Helmi, Mon. Not. R. Astron. Soc., \textbf{351}, 643 (2004)
\bibitem{vLeer1979} B. van Leer., J. of Comp. Phys., \textbf{32}, 101 (1979)
\bibitem{Wada2001} K. Wada, C.A. Norman., Astrophys. J., \textbf{547}, 172 (2001)
\bibitem{buma_khsa_Eremin-2010!VolGTU} M.A.~Eremin, A.V.~Khoperskov, S.A.~Khoperskov, Proceedings of Volgograd st. tech. univ., \textbf{13}, 24 (2010)
\bibitem{Khoperskov-etal-2010!z-str} A. Khoperskov,  D. Bizyaev,  N. Tyurina, M. Butenko,  Astron. Nachr., \textbf{331}, 731 (2010)
\bibitem{Khoperskov-2012!Halo-gas-spiral}
A.V.~Khoperskov, M.A.~Eremin, S.A.~Khoperskov et al., Astr. Report,  \textbf{56}, 16 (2012)
\bibitem{Khoperskov-Zasov-Tiurina-2003!GravitInstab}
A.V.~Khoperskov, A.V.~Zasov, N.V.~Tiurina, Astr. Report, \textbf{47},357 (2003)
\bibitem{Barnes1986} J. Barnes, P. Hut., Nature, \textbf{324}, 446 (1986)
\bibitem{Katz1992} N. Katz, Astrophys. J., \textbf{391}, 502 (1992)
\bibitem{Mori1997} M. Mori, Y. Yoshii, T. Tsujimoto et al., Astrophys. J., \textbf{478}, 21 (1997)
\bibitem{Rosen1995} A. Rosen, J.N Bregman, Astrophys. J., \textbf{440}, 634 (1995)
\bibitem{Navarro1993} J.F. Navarro,  S.D.M. White, Mon. Not. Roy. Astron. Soc., \textbf{265}, 271 (1993)
\bibitem{Wiebe1998} D.S.~Wiebe, A.V.~Tutukov, B.M.~Shustov, Astr. Report, \textbf{42}, 1 (1998)
\bibitem{Igumenshchev1990} I.V. Igumenshchev, B.M. Shustov, A.V. Tutukov, Astron. \& Astrophys.,\textbf{234}, 364 (1990)
\bibitem{Sellwood2011} J.A. Sellwood, Mon. Not. Roy. Astron. Soc., \textbf{410}, 1637 (2011)
\bibitem{Griv-2004!Jeans-unstable} E. Griv, M. Gedalin, Astron. J.,  \textbf{128}, 1965 (2004)
\bibitem{Khoperskov2010arxiv} A. Khoperskov ,  M. Eremin, S. Khoperskov et al., preprint (arXiv1007.2298) (2010)
\bibitem{Shustov2000} B.M. Shustov, D.S. Wiebe, Mon. Not. R. Astron. Soc. \textbf{319}, 1047 (2000)
\bibitem{Governato2012} F.Governato, A.Zolotov, A. Pontzen et al., preprint (arXiv202.0554v1) (2012)
\bibitem{Pontzen2012} A. Pontzen, F. Governato, preprint (arXiv:1106.0499v2) (2012)
\end{thebibliography}
\end{document}